\documentclass[twocolumn,preprintnumbers,amsmath,amssymb,prb,floatfix]{revtex4}

\usepackage{graphicx}
\usepackage{dcolumn}
\usepackage{bm}

\begin{document}


\title{Magnetic properties and domain structure of \\
(Ga,Mn)As films with perpendicular anisotropy}
\author{L. Thevenard} \email{laura.thevenard@lpn.cnrs.fr}
\author{L. Largeau}
\author{O. Mauguin}
\author{G. Patriarche}
\author{A. Lema\^{\i}tre}
 \affiliation{Laboratoire de Photonique et Nanostructures, CNRS, route de Nozay, F-91460, Marcoussis, France.}

\author{N. Vernier}
\author{J. Ferr\'e}

 \affiliation{Laboratoire de Physique des Solides,
UMR CNRS 8502, b\^at. 510, Universit\'e Paris-Sud, 91405 Orsay,
 France}

\date{\today}

\begin{abstract}

The ferromagnetism of a thin Ga$_{1-x}$Mn$_{x}$As layer with a
perpendicular easy anisotropy axis is investigated by means of
several techniques, that yield a consistent set of data on the
magnetic properties and the domain structure of this diluted
ferromagnetic semiconductor. The magnetic layer was grown under
tensile strain on a relaxed Ga$_{1-y}$In$_{y}$As buffer layer
using a procedure that limits the density of threading
dislocations. Magnetometry, magneto-transport and polar
magneto-optical Kerr effect (PMOKE) measurements reveal the high
quality of this layer, in particular through its high Curie
temperature (130~K) and well-defined magnetic anisotropy. We show
that magnetization reversal is initiated from a limited number of
nucleation centers and develops by easy domain wall propagation.
Furthermore, MOKE microscopy allowed us to characterize in detail
the magnetic domain structure. In particular we show that domain
shape and wall motion are very sensitive to some defects, which
prevents a periodic arrangement of the domains. We ascribed these
defects to threading dislocations emerging in the magnetic layer,
inherent to the growth mode on a relaxed buffer.
\end{abstract}

\maketitle

\section{\label{sec:level1} Introduction}

In the increasingly active field of research on spintronics,
impressive progress has been made in the understanding and
improvement of diluted magnetic semiconductors (DMS), in
particular of (Ga,Mn)As.\cite{Matsukura02} It is now well
established that ferromagnetism in this material stems from the
exchange interaction between the spins localized on the 3d shell
of the magnetic ions and the itinerant carriers.\cite{Dietl01a}
Such an interplay between spin-polarized holes and the randomly
distributed magnetic moments offers a novel set of phenomena not
found in usual ferromagnets, such as the strong dependence of the
Curie temperature on the carrier density ($p$), and the electrical
\cite{Ohno00} and optical \cite{Koshihara97} control of the
ferromagnetism. Another remarkable feature is the huge sensitivity
of the magnetic anisotropy on epitaxial strains, a behavior
related to the valence band anisotropy.\cite{Dietl01} Usually,
compressive strains favor an in-plane easy magnetization, as in
the case of a (GaMn)As layer grown on a GaAs substrate, although
this trend may be reversed at low carrier
density.\cite{Sawicki04,thevenard:182506} In counterpart, under
tensile strains, when (GaMn)As is grown on a (InGa)As buffer
layer,\cite{Shen97} the magnetization becomes spontaneously
oriented along the normal film to the film. Since only little
information is available so far on (GaMn)As films with
perpendicular magnetic anisotropy, we will focus here on their
magnetization reversal and domain study. Moreover, as already
reported for (GaMn)As with in-plane magnetic
anisotropy,\cite{Edmonds02} we show that the Curie temperature can
also be significantly enhanced after a convenient annealing
procedure of the sample that induces a higher metallic character
to the layer and more uniform properties, such as wider magnetic
domains.

Compared to the well-known magnetic behavior of 3d-metallic films,
there are still open questions on the incidence of the specific
nature of the exchange interaction in (GaMn)As, especially on the
dynamics of the magnetization reversal and related phenomena, such
as nucleation and domain wall motion. For in-plane magnetized
(Ga,Mn)As layers, large (micrometers-wide) and homogenous magnetic
domains have been observed by \citet{Welp03}  In that particular
case, magnetization reversal occurs by rare nucleations and
expansion of large $90^{\circ}$ and $180^{\circ}$ type of domains.

Up to now, only Scanning Hall Probe Microscopy has been used to
investigate the magnetic domain structure in (GaMn)As with
perpendicular anisotropy,\cite{Shono00,Pross04} but the area under
study was limited to a range of a few tens of square micrometers,
and information on the magnetic state was, in our opinion, too
limited to extract quantitative parameters. Nevertheless, a
typical domain width of a few micrometers has been
estimated,\cite{Shono00} consistently with theoretical arguments
assuming regular stripe-domains in such a magnetic
film.\cite{Dietl01a}

In this paper we report on the preparation, the characterization
and the magnetic properties of a thin (Ga,Mn)As layer grown in
tensile strain on an (In,Ga)As buffer. A thorough characterization
performed by combined X-ray diffraction, transmission electron
microscopy, magnetization, magneto-optics, conductivity and
magneto-transport measurements, confirmed the high structural and
magnetic quality of the sample, that is significantly improved
upon annealing.

Anomalous Hall effect and polar magneto-optical Kerr effect
(PMOKE) are suitable for accurate measurements of the
magnetization in thin ferromagnetic films with perpendicular
anisotropy. Both observables are related to the out-of-plane
component of the static magnetization M$_{\bot}$ through static
(Hall effect) or high frequency (PMOKE) conductivity non-diagonal
tensor elements. Also note that no diamagnetic contribution needs
to be deduced from Hall or PMOKE data since both effects are
insensitive to a possible spurious magnetism of the substrate. In
this article, PMOKE is used as the generic name. In fact, to be
less sensitive to the field-induced Faraday rotation in glass
windows of the cryostat, we measured the Polar Differential
Circular Reflectivity using left or right-handed circularly
polarized light. This effect is analogous, in reflection, to
magnetic circular dichroism in light transmission. Note that PMOKE
is much more sensitive to out-of-plane magnetization than
longitudinal MOKE for an in-plane magnetized
sample.\cite{Hrabovsky02,Moore03} PMOKE is sensitive to the
valence and conduction band splitting,\cite{Lang05,Beschoten99}
i.e. it varies like M$_{\bot}$, as soon as measurements are
carried far away from intense absorption bands, which is justified
here using green light wavelength ($\lambda=545$~nm). Moreover, we
checked that interference effects also have a negligible impact on
measurements. Using PMOKE, dynamics of the field-induced
magnetization reversal have been investigated here down to the
millisecond range. The magnetic relaxation has been further
analyzed by considering nucleation and domain expansion by wall
propagation. Our conclusions were confirmed from the direct
imaging of the magnetic domain structure by PMOKE microscopy.
Thus, micrometer-wide domains, pinned by defects, were evidenced
both in the remnant and in-field states.

\section{\label{sec:level2}Sample growth and structural characterization }

\begin{figure}
\includegraphics[scale=0.7]{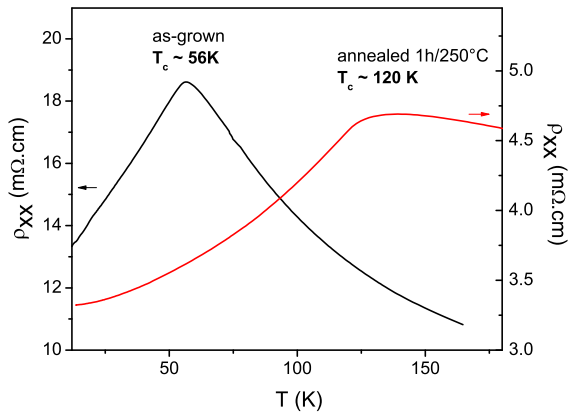}
\caption{\label{fig:Fig1}Thermal dependence of the longitudinal
resistivity for the as-grown (GaMn)As film (left), and for the
annealed sample (right).}
\end{figure}

The sample was prepared by molecular beam epitaxy. It is formed by
a 50 nm thick Ga$_{1-x}$Mn$_{x}$As layer grown on an
Ga$_{1-y}$In$_{y}$As relaxed buffer layer, itself deposited on a
semi-insulating (001) GaAs substrate. Special care was taken to
minimize the number of threading dislocations emerging in the
magnetic layer, following the technique developed by
 \citet{Harmand89} The Ga$_{1-y}$In$_{y}$As buffer consists first of a
 $\sim$ 0.5 $\mu$m thick layer grown by
increasing monotonously the In content from y = 0 to 9.8~\%. Then
2-3~$\mu$m of Ga$_{0.902}$In$_{0.098}$As were grown above before
depositing the Ga$_{1-x}$Mn$_{x}$As layer. The graded
Ga$_{1-y}$In$_{y}$As layer prevented the formation of a too large
amount of misfit dislocations, sources of threading dislocations
propagating along $\langle011\rangle$ towards the surface, as
usually observed in abrupt mismatched interfaces. Here,
dislocations were distributed along the graded layer, which limits
the nucleation of threading dislocations.\cite{Harmand89} Indeed,
a very low density of emerging dislocations, $(4 \pm 2).10^{4}$
cm$^{-2}$, was measured using an anisotropic revealing etchant.
Moreover, the substrate temperature was set to 400$^{\circ}$C
during the growth of the buffer to avoid the formation of
three-dimensional strain-induced islands, a process favored at
higher temperature.\cite{Harmand89} The Ga$_{1-x}$Mn$_{x}$As layer
was deposited at 250$^{\circ}$C. The surface is crossed-hatched,
due to surface roughness originating from bunches of misfit
dislocations propagating along $[110]$ and $[1\overline{1}0]$
inside the graded layer. $\langle004\rangle$ and
$\langle115\rangle$ X-ray diffraction space mappings showed that
the (In,Ga)As layer was almost completely relaxed (at 80\%),
insuring that the Ga$_{1-x}$Mn$_{x}$As layer was still under
tensile strain. The Mn concentration x$\sim$0.07 was determined by
comparison with Ga$_{1-x}$Mn$_{x}$As layers grown in the same
conditions, directly on GaAs (001). Finally, a part of the sample
was annealed under nitrogen atmosphere for 1 hour in a tube
furnace at 250$^{\circ}$C to out-diffuse the interstitial Mn
atoms, in order to improve magnetic properties.\cite{Edmonds02}
For magneto-transport measurements, the layer was processed into
Hall bars by UV-lithography and wet chemical etching. Ti/Au
contacts were used.

\begin{figure}
\includegraphics[bb=23 20 193 248,width=7.5 cm]{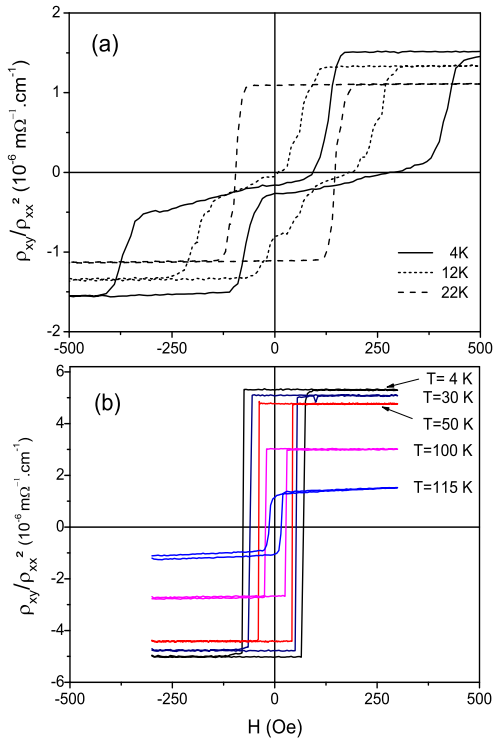}
\caption{\label{fig:Fig2}Magnetic hysteresis loops measured by
Hall resistivity at low temperatures: (a) as-grown film, (b)
annealed film. Sweeping rate of the magnetic field is 17~Oe/s.}
\end{figure}

\section{\label{sec:level3} Magnetic, magneto-transport and magneto-optic measurements }

\begin{figure}
\includegraphics[scale=0.8]{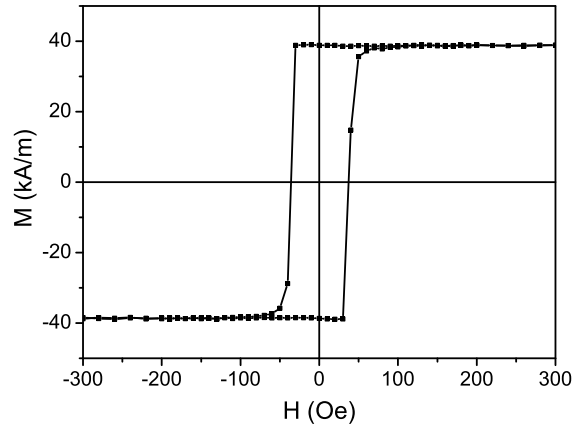}
\caption{\label{fig:Fig3}Hysteresis loop measured by SQUID
magnetometry at 4 K for the annealed sample. The magnetic field
was applied along the growth axis. }
\end{figure}

As quoted above, it is well known that the annealing process
improves both the transport and magnetic properties of (GaMn)As
deposited on (001) GaAs.\cite{Edmonds02} This is also true for a
tensile-strained (GaMn)As film grown on (In,Ga)As, as checked by
resistivity (Fig. \ref{fig:Fig1}), and Hall hysteresis loop (Fig.
\ref{fig:Fig2}) extracted from the transverse resistivity, as
described later in the text. The resistivity of the as-grown
Ga$_{0.93}$Mn$_{0.07}$As film on (In,Ga)As is rather high
($\rho_{xx} =13$~mW.cm at 4K), showing a maximum at T$_{c}$ =~56 K
(Fig. \ref{fig:Fig1}). In counterpart, the annealed film exhibits
a much lower resistivity, i.e. a more metallic character
($\rho_{xx}$ =~2.58 mW.cm at 4K) at low temperature, and a
remarkably higher Curie temperature, $T_{c}\sim120-130$~K. As
often found for as-grown samples with in-plane magnetic
anisotropy, the Hall hysteresis loops of the non-annealed sample
exhibit a complex shape at low temperature \cite{Lang05}. In the
present case, and coherently with previous
results,\cite{Sawicki04,liu:063904} the evolution of the loop
shape between 22 K to 4 K (Fig. \ref{fig:Fig2}a) suggests an
out-of-plane to in-plane spin reorientation transition when
decreasing the temperature. In counterpart, the hysteresis loops
of the annealed Ga$_{0.93}$Mn$_{0.07}$As film (Fig.
\ref{fig:Fig2}b) are perfectly square at all temperatures below
T$_{c}$, consistently with a marked out-of-plane magnetic
anisotropy. The magnetization of the annealed sample, measured by
a Superconducting Quantum Interference Device (SQUID) also shows a
behavior very consistent with Hall measurements, namely a square
hysteresis loop at 4 K (Fig. \ref{fig:Fig3}), with a coercive
field of 40 $\pm$ 5~Oe, obtained with a low field sweeping rate
(0.08~Oe/s). In the following, we will only focus on properties of
the annealed film.

\begin{figure}
\includegraphics[scale=0.7]{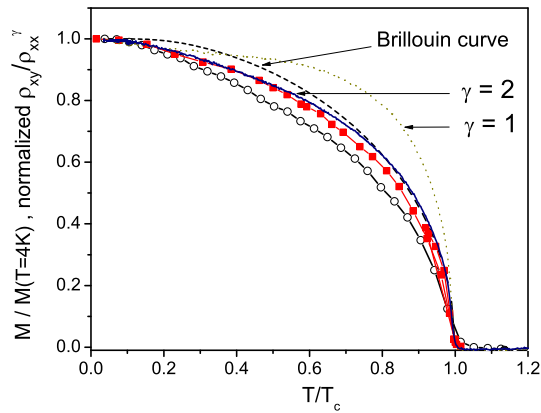}
\caption{\label{fig:Fig4}Temperature dependence of the
magnetization (open circles) of the annealed sample measured by
SQUID magnetometry under $H =250$~Oe applied along the growth
axis, compared to that of the Hall effect
$\rho_{xy}/\rho_{xx}^{\gamma}$ with $\gamma$ = 1 (dotted line) or
$\gamma$ =~2 (full line), and of the remnant PMOKE signal (closed
squares).  For each curve, the data were normalized to their low
temperature value, and plotted as a function of the reduced
temperature $T/T_{c}$. For comparison, the mean-field $S = 5/2$
Brillouin function is also plotted (dashed line).}
\end{figure}

The temperature dependence of the magnetization measured by SQUID
in a small magnetic field (250~Oe) applied along the growth axis
is presented in Fig. \ref{fig:Fig4}. It shows a smooth decline
with increasing temperature and magnetization once again cancels
at $T_{c} = 130$~K. For comparison we have plotted the mean-field
Brillouin function expected for an ideal $S = 5/2$ ferromagnet.
Only a small departure from the Brillouin curve is observed. Such
a behavior is predicted in GaMnAs layers,\cite{Dietl01} and may
arise partly from the fact that the spin-splitting of the valence
band is comparable to the Fermi energy, as explained within the
mean-field theory frame. We note that the discrepancy between
these two curves remains fairly low, reaching 20 \% at most,
giving another indication of the high quality of this sample.
SQUID and PMOKE measurements also prove that the annealed (GaMn)As
film is not ferromagnetic at room temperature. More precisely, as
compared to results for MnAs clusters,\cite{Primus2005} high field
(6.5~kOe) PMOKE data exclude that more than 0.5 \% of Mn ions
enter in the formation of ferromagnetic MnAs nanoclusters.

We now return  to the Hall measurements. In GaMnAs layers, the
anomalous Hall resistivity, $\rho_{xy}^{a}$, has been found to be
proportional to $M_{\bot}$, so that $\rho_{xy}^{a} =
R_{a}(\rho_{xx})M_{\bot}$, R$_{a}$ being called the anomalous Hall
term. It directly probes the carrier-mediated ferromagnetism,
since it is induced, through the spin-orbit coupling, by
anisotropic scattering of spin-up and spin-down carriers. R$_{a}$
was shown to be proportional to $\rho_{xx}^{\gamma}$, where
$\rho_{xx}$ is the sheet resistivity (Fig. \ref{fig:Fig1}). The
value of $\gamma$  is still a matter of debate (for a review, see
\citet{Sinova04b}). It was first proposed to depend on the spin
scattering mechanism, with $\gamma$ =~1 in the case of "skew
scattering", or $\gamma$ =~2 for "side-jump
scattering".\cite{Berger79} More recently, \citet{Jungwirth02}
showed that the dominant contribution to R$_{a}$ would be due to a
scattering-independent topological contribution, which also gives
$\gamma$ =~2.

Therefore, the anomalous Hall resistivity, gives access to
perpendicular hysteresis loops, yielding information for example
on the coercivity (as seen in Fig. \ref{fig:Fig2} where the ratio
$\rho_{xy}/\rho_{xx}^{\gamma}$ was plotted using $\gamma$ =~2),
the remnant magnetization, and the thermal dependence of the
saturated magnetization.

The thermal dependence of the remnant Hall effect is also
presented in Fig. \ref{fig:Fig4} as a function of $T/T_{c}$,
considering $\gamma$ =~1 or $\gamma$ =~2. The choice of $\gamma$
has a rather large impact on the result below $T_{c}$. For
$\gamma$ = 1, the Hall effect is nearly constant from $T = 0$~K up
to 0.6 $T_{c}$, but shows a rapid drop at T$_{c}$, whereas for
$\gamma$ =~2, it decreases more progressively at high temperature.
Both curves are always located above the magnetization curve up to
T$_{c}$. While the thermal dependence of the $\gamma$ = 2 curve is
similar to that of the magnetization within 10\% at most, the
$\gamma$ =~1 curve exceeds magnetization values by up to 36\%.
Furthermore the $\gamma$ =~2 curve closely follows the Brillouin
curve. Lastly, the $\gamma$ =~2 curve matches entirely the remnant
PMOKE signal (Fig. \ref{fig:Fig4}), which is another measure of
the magnetization. Note that the anomalous Hall term R$_{a}$ may
not remain perfectly constant within the whole temperature range.
Nevertheless, we obtained a very consistent set of results when
comparing the ratio $\rho_{xy}$/$\rho_{xx}^{\gamma}$ with $\gamma$
=~2 to the magnetization and PMOKE data.

\begin{figure}
\includegraphics[scale=0.4]{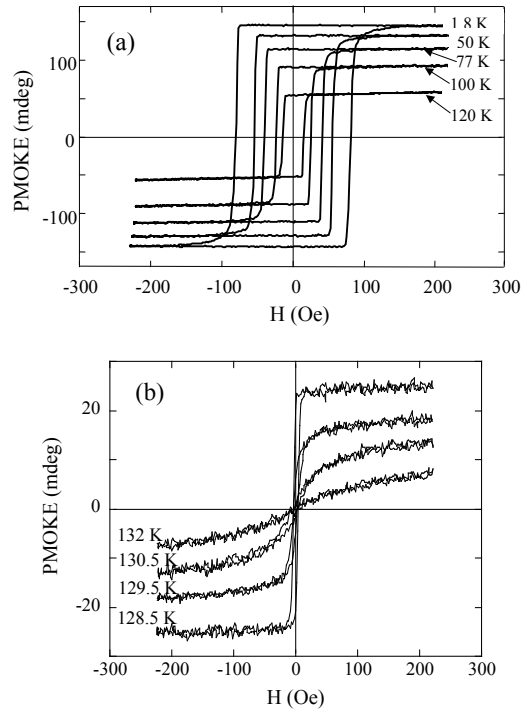}
\caption{\label{fig:Fig5}a) PMOKE hysteresis loops measured at
different temperatures for the annealed (GaMn)As film. (b) in the
vicinity of $T_c$ = 130 K.}
\end{figure}

As in Hall magnetometry, we measured very square PMOKE magnetic
hysteresis loops over the full range of temperatures up to T$_{c}$
=~130 K (Fig. \ref{fig:Fig5}a). This proves that domain reversal
occurs very abruptly and uniformly over the sample, as soon as one
reaches the coercive field. The quality of the
Ga$_{0.93}$Mn$_{0.07}$As annealed film is also confirmed by the
abrupt disappearance of the remnant magnetization at $T_c$ (Fig.
\ref{fig:Fig5}b). At low temperature, loops do not saturate in
field as abruptly in PMOKE (Fig. \ref{fig:Fig5}a) as in Hall
magnetometry (Fig. \ref{fig:Fig2}). This is certainly due to the
fact that the probed region is larger in PMOKE (1 mm$^{2}$) than
in Hall magnetometry (10$^{-2}$ mm$^{2}$). As we shall see in
PMOKE microscopy, the occurrence of hysteresis loop tails is due
to a few hard defects that pin small reversed magnetized
filamentary domains even in fields larger than the coercive field.
At finite temperature, the coercive field deduced from PMOKE
measurements is always expected to be larger than in Hall
magnetometry and magnetization measurements because of the much
faster field sweeping rate in the former case
\cite{Moore03,Ferre2001} (200 Oe/s).

\section{\label{sec:level4} Magnetization reversal dynamics and
domains}

As mentioned above, dynamics of the magnetization reversal
manifests itself by an increase of the coercive field with the
field sweeping rate. A clearer view of the magnetization reversal
dynamic behavior is given from magnetic relaxation curves, also
called magnetic after-effect. The film is first magnetically
saturated in a positive field of 135 Oe, and suddenly inverted, at
t = 0, to a negative value $- H$, that is smaller than the
measured coercive field. The magnetization of the film being in a
metastable state, it tends to reverse gradually with time towards
the negative saturated magnetization value. Obviously, the
relaxation of the magnetization becomes faster when increasing
$H$. Previous results have been reported on (GaMn)As films with
in-plane anisotropy.\cite{Moore03}

\begin{figure}
\includegraphics[bb=79 50 502 734,width=7.5 cm]{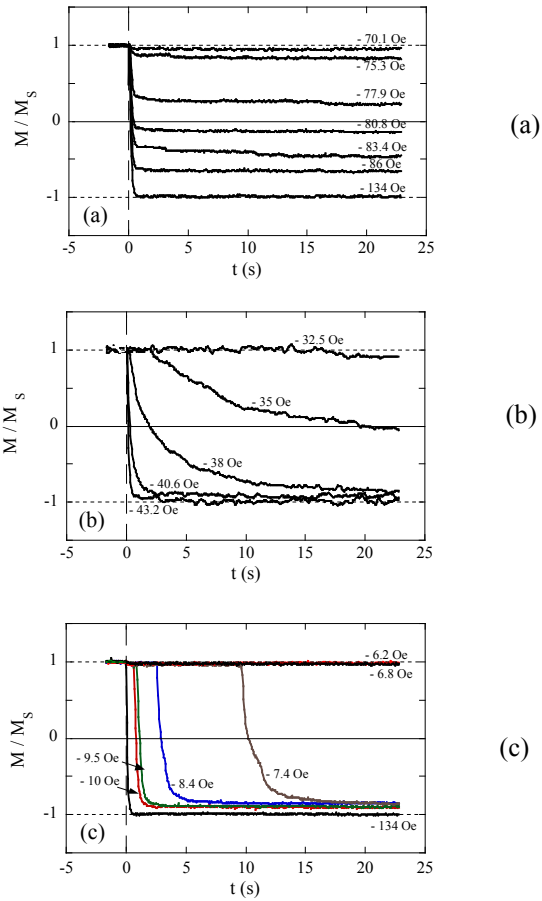}
\caption{\label{fig:Fig6}: Magnetic after-effect relaxation curves
for the annealed GaMnAs film and for different field values.
(a)~$T = 1.8$~K, (b)~$T = 77$~K, (c)~$T = 113.5$~K.}
\end{figure}

\begin{figure}
\includegraphics[bb=68 216 540 574,scale=0.25]{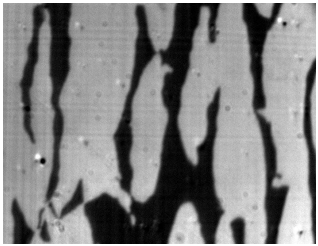}
\caption{\label{fig:Fig7}AC-demagnetized state of the annealed
(GaMn)As film at 77 K. Up (down)-magnetized domains are appearing
in black (white). Image size : 135 $\mu$m x 176 $\mu$m.}
\end{figure}

\begin{figure*}
\includegraphics[width=0.8 \textwidth,bb=68 216 540 574]{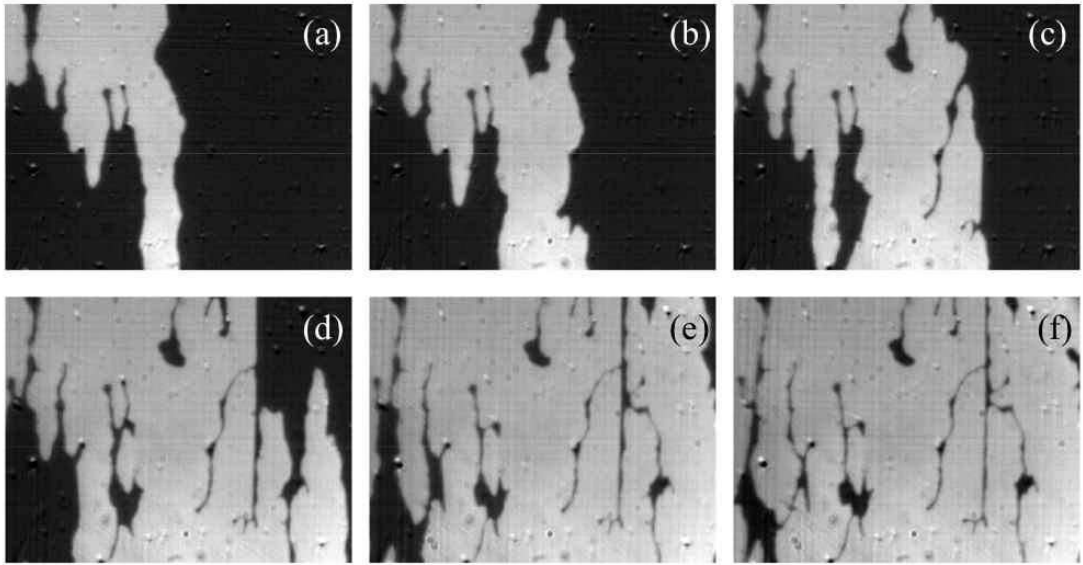}
\caption{\label{fig:Fig8}Successive snapshots of the magnetization
reversal at 77 K, observed by PMOKE microscopy, after having
applied a field of $H = - 20.5$~Oe during one minute. The lag time
between successive images is: 10s between (a) and (b), and (b) and
(c). It is 30s, 2 minutes, and 15 minutes between (c) and (d), (d)
and (e), (e) and (f), respectively. The image size is 135 $\mu$m x
176 $\mu$m.}
\end{figure*}

Results on the magnetic after-effect for the annealed (GaMn)As
film, measured by PMOKE at 1.8 K, 77K, and 100 K, i.e. below
$T_{c} = 130$~K, are depicted on Fig. \ref{fig:Fig6} a,b,c. The
time dependent variation of the PMOKE signal is measured on a film
area delimited by the light spot size (about 1 mm$^{2}$).  As we
shall see later, the magnetization reversal is initiated by rare
nucleation events, and develops by subsequent domain wall
propagation. At very low temperature, the thermal activation is
not very efficient, so that the applied field H must overcome the
height of local energy barriers for wall propagation. In other
words, if H is higher than the nucleation field and most of the
distributed pinning barriers, the wall will propagate rapidly
until reaching centers with high energy barriers. After that,
walls will only move by poorly efficient thermal activation. This
behavior is well revealed on relaxation curves at T =~1.8 K (Fig.
\ref{fig:Fig6}a). At temperatures closer to T$_{c}$, most of the
propagation energy barriers become weaker than the energy barrier
for nucleation. Thus, a thermally activated lag time is required
for nucleation; it is followed by rapid domain wall propagation.
This gives rise to typical magnetization relaxation curves
presented on Fig. \ref{fig:Fig6}c, at 113.5 K, when nucleation
occurs outside of the investigated light spot area. As expected,
the probability of nucleation, which follows a thermally activated
Arrhenius law,\cite{Ferre2001} increases rapidly with the applied
field. In the intermediate temperature range ($T =77$~K), one
generally observes a rather monotonous decrease of the
magnetization (Fig. \ref{fig:Fig6}b) when the light spot only
checks a homogeneous sample area. As we shall see later, the
investigated region can contain several strong pinning centers. In
that case, the domain wall propagation is suddenly slowed down by
local pinning, and then accelerated through a depinning process
involving the relaxation of the bending energy. Similar dynamic
behavior has been already evidenced in a pure metallic Au/Co/Au
ultrathin film structure.\cite{Ferre2001}

Finally, the domain structure and dynamics have been investigated
by PMOKE microscopy. A similar set-up, working at room
temperature, has been previously
described.\cite{A.Hubert1198,Ferre2001} Since the long distance
objective was located outside the cryostat vessel, the optical
resolution was limited here to about 1.5~$\mu$m at the used red
optical wavelength ($\lambda$ =~650 nm). It is well known that the
equilibrium demagnetized state of a perfect film with
perpendicular anisotropy must be a stripe domain structure from
which microscopic parameters may be deduced.\cite{A.Hubert1198}
However, as soon as extrinsic defects are efficient enough to pin
the walls, the demagnetized state becomes highly perturbed and
tends to decorate the assembly of pinning centers. As depicted on
Fig. \ref{fig:Fig7}, this is the case of our
Ga$_{0.93}$Mn$_{0.07}$As annealed film, where large up and
down-magnetized (typically 15~$\mu$m wide) domains are stabilized
in the demagnetized state without reminiscence of any periodic
ribbon domain structure that is expected in a defect-free sample.

\begin{figure}
\includegraphics[bb=43 180 552 591, scale=0.35]{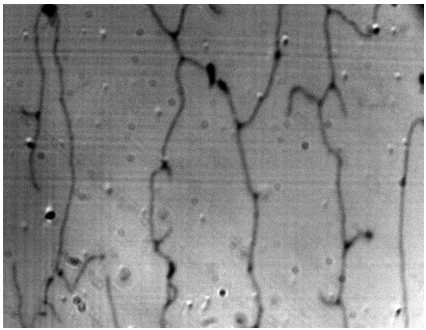}
\caption{\label{fig:Fig9}Remnant domain structure after applying a
field of $H = - 50$~Oe, and switching it to zero. Image size : 135
$\mu$m x 176 $\mu$m.  Stable non-reversed filamentary domains
(360$^{\circ}$ domain walls) are clearly observed.}
\end{figure}

\begin{figure}
\includegraphics[bb=56 203 542 677,width=7.5 cm]{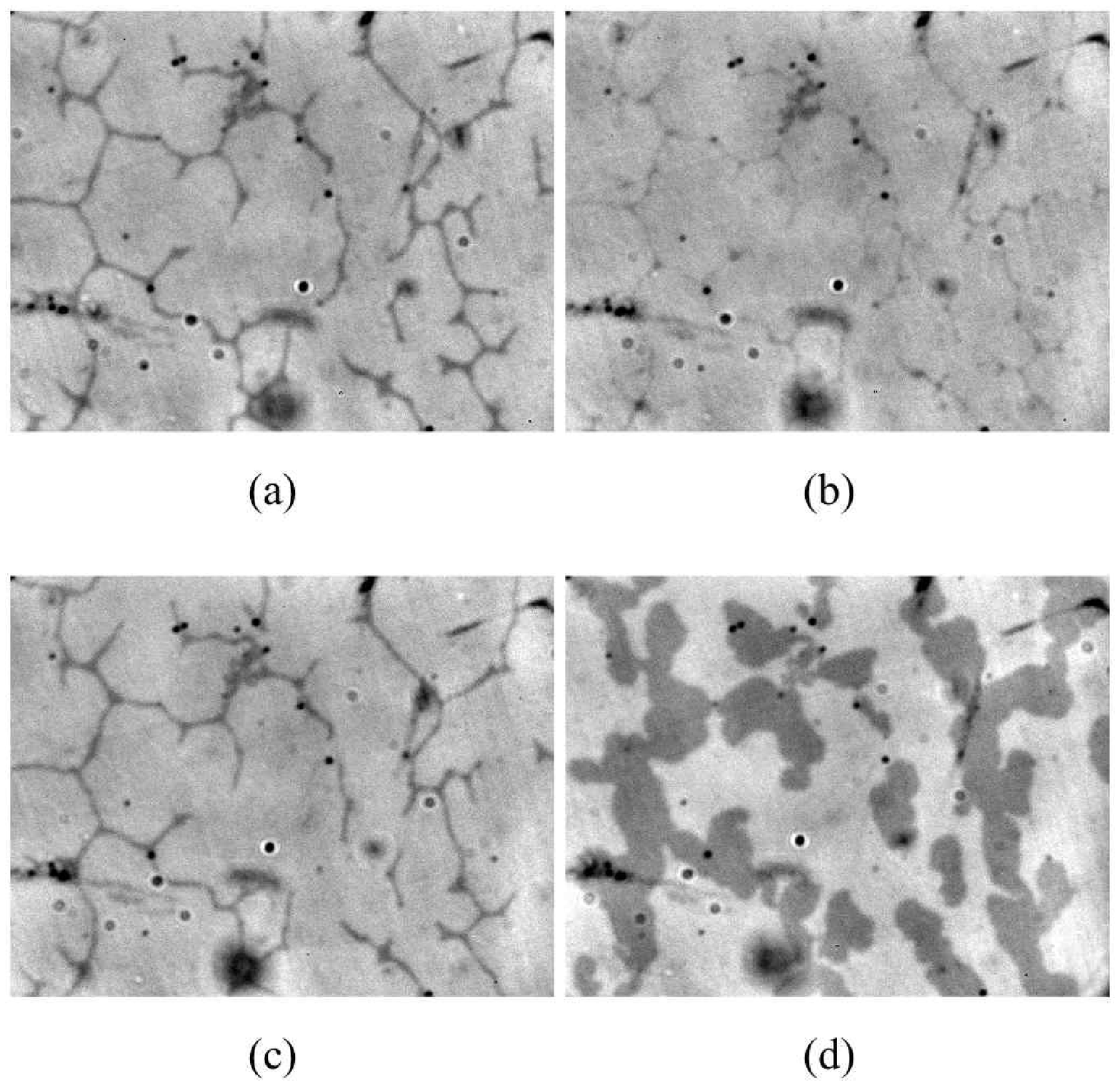}
\caption{\label{fig:Fig10}Successive PMOKE snapshots of the
magnetic domain structure on another part of the annealed (GaMn)As
film. (a) The sample is first saturated in a positive field (black
up-magnetized domains), and subsequently submitted to a negative
field $H=-140$~Oe  for a few seconds, and switched to zero.
Up-magnetized filamentary domains are frozen in the remnant state.
(b) A negative field ($H=-112$~Oe)  is applied. The filamentary
structure is shrinked, since the field tends to reduce their
width. (c) Switching again the field back to zero, a remnant state
similar to the previous one (a), is observed, except for a few
branches that have disappeared under field due to their depinning
on defects. (d) Finally, a positive field ($H=53$~Oe), close to
the coercive value, has been applied during 5s. The snapshot is
obtained after switching off the field. The image size is 135
$\mu$m x 176 $\mu$m.}
\end{figure}

To get a better understanding of the magnetization reversal
process, snapshots of successive magnetic states of an annealed
Ga$_{0.93}$Mn$_{0.07}$As film, have been recorded at 77~K, using a
magnetic after-effect procedure (Fig. \ref{fig:Fig8}). After
saturating the magnetization in a positive field, a weak enough
field (- 20.5~Oe) was applied to produce a slow magnetization
reversal. An order of magnitude of the domain wall velocity is 10
$\mu$m/s in a field of $H = - 12$~Oe. After nucleation of a
down-magnetized nucleation center, located above the image frame,
a down-magnetized (white) domain invades progressively the
visualized area at the expense of the up-magnetized state (black)
(Fig. \ref{fig:Fig8}a to d). So, nucleation is a rare event, and
the magnetization reversal occurs by a rather uniform wall
propagation. The front of propagation is irregular since walls are
often pinned by defects that slow down their motion. After
skirting a defect, the domain continues to grow by leaving a
stable up-magnetized filamentary domain, also called unwinding
360$^{\circ}$ wall, along the direction of motion. \cite{Chappert}
During this process, random magnetization jumps, due to successive
local depinnings, can take place after long lag times
(Fig.\ref{fig:Fig8}d to f); the resulting magnetization reversal
slows down progressively. Even after waiting a long time, a field
of - 50~Oe, i.e. equal to the coercive field, is not strong enough
to reverse the magnetization inside 360$^{\circ}$ walls (Fig.
\ref{fig:Fig9}). This common property for 360$^{\circ}$ walls in
defected metallic ferromagnetic films \cite{Chappert} explains
well the presence of hysteresis loop tails. From these snapshots
(Fig. \ref{fig:Fig8}), we can estimate an average number of strong
pinning centers of about 10$^{5}$ centers/cm$^{2}$ over the
considered image area; this number is rather low, but comparable
to the density of emerging dislocations deduced previously.

To confirm the role of emerging dislocations, we examined the
effect of a field on a vestigial filamentary 360$^{\circ}$ wall
structure, such as that shown in Fig. \ref{fig:Fig9}. The zero
field remnant magnetic pattern obtained after applying a rather
large negative field, $H=-140$~Oe, but during a short time,
depicts an up-magnetized filamentary state. Defects are localized
at the end and at the intersection between branches of the
filamentary state (Fig. \ref{fig:Fig10}a). After applying once
again a negative field, $H=-112$~Oe (Fig. \ref{fig:Fig10}b),
branches become far thinner, due to Zeeman forces that partly
compensate the dipolar repulsive force between facing walls in the
filamentary domains. After switching off the field again to return
to a remnant state, the thickness of the new filamentary domains
is restored, but a very limited number of up-magnetized branches
disappear (compare Fig. \ref{fig:Fig10} a and c). The application
of a rather weak positive field ($H=52$~Oe) blows up the
up-magnetized filamentary domains, but preserves bottlenecks at
defect positions. This means that the filamentary, or
360$^{\circ}$ domain wall structure, is a highly stable state that
is pinned by a small number of defects. Thus, this vestigial
filamentary domain structure is clearly initiated and stabilized
by point defects, most probably emerging dislocations.

\section{\label{sec:level5} Conclusion}

Magnetization, magneto-transport and PMOKE measurements bring
together very consistent data, and enable us to conclude that,
albeit the growth on a relaxed (InGa)As buffer, the annealed
sample is of high quality, both from the magnetic and transport
points of view. These results confirm that (GaMn)As films with
perpendicular anisotropy can be grown on a relaxed buffer with
nearly the same quality as samples with in-plane anisotropy, grown
directly on GaAs. Such good properties allowed to implement PMOKE
experiments in optimal conditions for magnetic dynamical and
imaging studies. In particular, the magnetization reversal
process, which occurs via domain nucleation at rare places
followed by fast and quasi-isotropic domain wall propagation, has
been observed and interpreted. In spite of the profound difference
in the origin of magnetism in metallic and DMS films,
micromagnetism and magnetization reversal dynamics show obvious
similarities. Emerging dislocations, inherent to growth on a
relaxed buffer, form pinning centers for domain walls during the
magnetization reversal process at low temperature, giving rise to
vestigial filamentary domains, or 360$^{\circ}$ walls.
Nevertheless, their low density might allow to fabricate devices,
for example to built tunnel junctions with large magnetoresistance
or to study current-induced domain-wall propagation in
sub-micrometric stripe structures.\cite{Yamanouchi04}

\begin{acknowledgments}
This work has been supported by the R\'{e}gion Ile de France, the
Conseil G\'{e}n\'{e}ral de l'Essonne and through the Action
Concert\'{e}e Incitative BOITQUAN and ANR PNano MOMES. We would
like to thank J.-C. Harmand, J.-P. Jamet, and A. Mougin for very
fruitful discussions, as well as P. Monod and I. Robert for giving
us access to their equipment.
\end{acknowledgments}


\begin{thebibliography}{26}
\expandafter\ifx\csname
natexlab\endcsname\relax\def\natexlab#1{#1}\fi
\expandafter\ifx\csname bibnamefont\endcsname\relax
  \def\bibnamefont#1{#1}\fi
\expandafter\ifx\csname bibfnamefont\endcsname\relax
  \def\bibfnamefont#1{#1}\fi
\expandafter\ifx\csname citenamefont\endcsname\relax
  \def\citenamefont#1{#1}\fi
\expandafter\ifx\csname url\endcsname\relax
  \def\url#1{\texttt{#1}}\fi
\expandafter\ifx\csname
urlprefix\endcsname\relax\def\urlprefix{URL }\fi
\providecommand{\bibinfo}[2]{#2}
\providecommand{\eprint}[2][]{\url{#2}}

\bibitem[{\citenamefont{Matsukura et~al.}(2002)\citenamefont{Matsukura, Ohno,
  and Dietl}}]{Matsukura02}
\bibinfo{author}{\bibfnamefont{F.}~\bibnamefont{Matsukura}},
  \bibinfo{author}{\bibfnamefont{H.}~\bibnamefont{Ohno}}, \bibnamefont{and}
  \bibinfo{author}{\bibfnamefont{T.}~\bibnamefont{Dietl}},
  \emph{\bibinfo{title}{Handbook of Magnetic Materials}}
  (\bibinfo{publisher}{Elsevier}, \bibinfo{address}{Amsterdam},
  \bibinfo{year}{2002}), vol.~\bibinfo{volume}{14}, p.~\bibinfo{pages}{1}.

\bibitem[{\citenamefont{Dietl et~al.}(2001{\natexlab{a}})\citenamefont{Dietl,
  Ohno, and Matsukura}}]{Dietl01a}
\bibinfo{author}{\bibfnamefont{T.}~\bibnamefont{Dietl}},
  \bibinfo{author}{\bibfnamefont{H.}~\bibnamefont{Ohno}}, \bibnamefont{and}
  \bibinfo{author}{\bibfnamefont{F.}~\bibnamefont{Matsukura}},
  \bibinfo{journal}{Phys. Rev. B}
  \textbf{\bibinfo{volume}{63}}, \bibinfo{eid}{195205}
  (\bibinfo{year}{2001}{\natexlab{a}}).

\bibitem[{\citenamefont{Ohno et~al.}(2000)\citenamefont{Ohno, Chiba, Matsukura,
  Omiya, Abe, Dietl, Ohno, and Ohtani}}]{Ohno00}
\bibinfo{author}{\bibfnamefont{H.}~\bibnamefont{Ohno}},
  \bibinfo{author}{\bibfnamefont{D.}~\bibnamefont{Chiba}},
  \bibinfo{author}{\bibfnamefont{F.}~\bibnamefont{Matsukura}},
  \bibinfo{author}{\bibfnamefont{T.}~\bibnamefont{Omiya}},
  \bibinfo{author}{\bibfnamefont{E.}~\bibnamefont{Abe}},
  \bibinfo{author}{\bibfnamefont{T.}~\bibnamefont{Dietl}},
  \bibinfo{author}{\bibfnamefont{Y.}~\bibnamefont{Ohno}}, \bibnamefont{and}
  \bibinfo{author}{\bibfnamefont{K.}~\bibnamefont{Ohtani}},
  \bibinfo{journal}{Nature} \textbf{\bibinfo{volume}{408}},
  \bibinfo{pages}{944} (\bibinfo{year}{2000}).

\bibitem[{\citenamefont{Koshihara et~al.}(1997)\citenamefont{Koshihara, Oiwa,
  Hirasawa, Katsumoto, Iye, Urano, Takagi, and Munekata}}]{Koshihara97}
\bibinfo{author}{\bibfnamefont{S.}~\bibnamefont{Koshihara}},
  \bibinfo{author}{\bibfnamefont{A.}~\bibnamefont{Oiwa}},
  \bibinfo{author}{\bibfnamefont{M.}~\bibnamefont{Hirasawa}},
  \bibinfo{author}{\bibfnamefont{S.}~\bibnamefont{Katsumoto}},
  \bibinfo{author}{\bibfnamefont{Y.}~\bibnamefont{Iye}},
  \bibinfo{author}{\bibfnamefont{C.}~\bibnamefont{Urano}},
  \bibinfo{author}{\bibfnamefont{H.}~\bibnamefont{Takagi}}, \bibnamefont{and}
  \bibinfo{author}{\bibfnamefont{H.}~\bibnamefont{Munekata}},
  \bibinfo{journal}{Phys. Rev. Lett.}
  \textbf{\bibinfo{volume}{78}}, \bibinfo{pages}{4617}
  (\bibinfo{year}{1997}).

\bibitem[{\citenamefont{Dietl et~al.}(2001{\natexlab{b}})\citenamefont{Dietl,
  K\"{o}nig, and MacDonald}}]{Dietl01}
\bibinfo{author}{\bibfnamefont{T.}~\bibnamefont{Dietl}},
  \bibinfo{author}{\bibfnamefont{J.}~\bibnamefont{K\"{o}nig}},
  \bibnamefont{and} \bibinfo{author}{\bibfnamefont{A.~H.}
  \bibnamefont{MacDonald}}, \bibinfo{journal}{Phys. Rev. B}
  \textbf{\bibinfo{volume}{64}}, \bibinfo{pages}{241201(R)} (\bibinfo{year}{2001}{\natexlab{b}}).

\bibitem[{\citenamefont{Sawicki et~al.}(2004)\citenamefont{Sawicki, Matsukura,
  Idziaszek, Dietl, Schott, Ruester, Gould, Karczewski, Schmidt, and
  Molenkamp}}]{Sawicki04}
\bibinfo{author}{\bibfnamefont{M.}~\bibnamefont{Sawicki}},
  \bibinfo{author}{\bibfnamefont{F.}~\bibnamefont{Matsukura}},
  \bibinfo{author}{\bibfnamefont{A.}~\bibnamefont{Idziaszek}},
  \bibinfo{author}{\bibfnamefont{T.}~\bibnamefont{Dietl}},
  \bibinfo{author}{\bibfnamefont{G.~M.} \bibnamefont{Schott}},
  \bibinfo{author}{\bibfnamefont{C.}~\bibnamefont{Ruester}},
  \bibinfo{author}{\bibfnamefont{C.}~\bibnamefont{Gould}},
  \bibinfo{author}{\bibfnamefont{G.}~\bibnamefont{Karczewski}},
  \bibinfo{author}{\bibfnamefont{G.}~\bibnamefont{Schmidt}}, \bibnamefont{and}
  \bibinfo{author}{\bibfnamefont{L.~W.} \bibnamefont{Molenkamp}},
  \bibinfo{journal}{Phys. Rev. B}
  \textbf{\bibinfo{volume}{70}}, \bibinfo{pages}{245325(R)}
   (\bibinfo{year}{2004}).

\bibitem[{\citenamefont{Thevenard et~al.}(2005)\citenamefont{Thevenard,
  Largeau, Mauguin, Lemaitre, and Theys}}]{thevenard:182506}
\bibinfo{author}{\bibfnamefont{L.}~\bibnamefont{Thevenard}},
  \bibinfo{author}{\bibfnamefont{L.}~\bibnamefont{Largeau}},
  \bibinfo{author}{\bibfnamefont{O.}~\bibnamefont{Mauguin}},
  \bibinfo{author}{\bibfnamefont{A.}~\bibnamefont{Lemaitre}}, \bibnamefont{and}
  \bibinfo{author}{\bibfnamefont{B.}~\bibnamefont{Theys}},
  \bibinfo{journal}{Appl. Phys. Lett.} \textbf{\bibinfo{volume}{87}},
  \bibinfo{eid}{182506} (\bibinfo{year}{2005}).

\bibitem[{\citenamefont{Shen et~al.}(1997)\citenamefont{Shen, Ohno, Matsukura,
  Sugawara, Akiba, Kuroiwa, Oiwa, Endo, Katsumoto, and Iye}}]{Shen97}
\bibinfo{author}{\bibfnamefont{A.}~\bibnamefont{Shen}},
  \bibinfo{author}{\bibfnamefont{H.}~\bibnamefont{Ohno}},
  \bibinfo{author}{\bibfnamefont{F.}~\bibnamefont{Matsukura}},
  \bibinfo{author}{\bibfnamefont{Y.}~\bibnamefont{Sugawara}},
  \bibinfo{author}{\bibfnamefont{N.}~\bibnamefont{Akiba}},
  \bibinfo{author}{\bibfnamefont{T.}~\bibnamefont{Kuroiwa}},
  \bibinfo{author}{\bibfnamefont{A.}~\bibnamefont{Oiwa}},
  \bibinfo{author}{\bibfnamefont{A.}~\bibnamefont{Endo}},
  \bibinfo{author}{\bibfnamefont{S.}~\bibnamefont{Katsumoto}},
  \bibnamefont{and} \bibinfo{author}{\bibfnamefont{Y.}~\bibnamefont{Iye}},
  \bibinfo{journal}{J. Cryst. Growth} \textbf{\bibinfo{volume}{175-176}},
  \bibinfo{pages}{1069} (\bibinfo{year}{1997}).

\bibitem[{\citenamefont{Edmonds et~al.}(2002)\citenamefont{Edmonds, Wang,
  Campion, Neumann, Foxon, Gallagher, and Main}}]{Edmonds02}
\bibinfo{author}{\bibfnamefont{K.~W.} \bibnamefont{Edmonds}},
  \bibinfo{author}{\bibfnamefont{K.~Y.} \bibnamefont{Wang}},
  \bibinfo{author}{\bibfnamefont{R.~P.} \bibnamefont{Campion}},
  \bibinfo{author}{\bibfnamefont{A.~C.} \bibnamefont{Neumann}},
  \bibinfo{author}{\bibfnamefont{C.~T.} \bibnamefont{Foxon}},
  \bibinfo{author}{\bibfnamefont{B.~L.} \bibnamefont{Gallagher}},
  \bibnamefont{and} \bibinfo{author}{\bibfnamefont{P.~C.} \bibnamefont{Main}},
  \bibinfo{journal}{Appl. Phys. Lett.} \textbf{\bibinfo{volume}{81}},
  \bibinfo{pages}{3010} (\bibinfo{year}{2002}).

\bibitem[{\citenamefont{Welp et~al.}(2003)\citenamefont{Welp, Vlasko-Vlasov,
  Liu, Furdyna, and Wojtowicz}}]{Welp03}
\bibinfo{author}{\bibfnamefont{U.}~\bibnamefont{Welp}},
  \bibinfo{author}{\bibfnamefont{V.~K.} \bibnamefont{Vlasko-Vlasov}},
  \bibinfo{author}{\bibfnamefont{X.}~\bibnamefont{Liu}},
  \bibinfo{author}{\bibfnamefont{J.~K.} \bibnamefont{Furdyna}},
  \bibnamefont{and}
  \bibinfo{author}{\bibfnamefont{T.}~\bibnamefont{Wojtowicz}},
  \bibinfo{journal}{Phys. Rev. Lett.} \textbf{\bibinfo{volume}{90}},
  \bibinfo{eid}{167206} (\bibinfo{year}{2003}).

\bibitem[{\citenamefont{Shono et~al.}(2000)\citenamefont{Shono, Hasegawa,
  Fukumura, Matsukura, and Ohno}}]{Shono00}
\bibinfo{author}{\bibfnamefont{T.}~\bibnamefont{Shono}},
  \bibinfo{author}{\bibfnamefont{T.}~\bibnamefont{Hasegawa}},
  \bibinfo{author}{\bibfnamefont{T.}~\bibnamefont{Fukumura}},
  \bibinfo{author}{\bibfnamefont{F.}~\bibnamefont{Matsukura}},
  \bibnamefont{and} \bibinfo{author}{\bibfnamefont{H.}~\bibnamefont{Ohno}},
  \bibinfo{journal}{Appl. Phys. Lett.} \textbf{\bibinfo{volume}{77}},
  \bibinfo{pages}{1363} (\bibinfo{year}{2000}).

\bibitem[{\citenamefont{Pross et~al.}(2004)\citenamefont{Pross, Bending,
  Edmonds, Campion, Foxon, and Gallagher}}]{Pross04}
\bibinfo{author}{\bibfnamefont{A.}~\bibnamefont{Pross}},
  \bibinfo{author}{\bibfnamefont{S.}~\bibnamefont{Bending}},
  \bibinfo{author}{\bibfnamefont{K.}~\bibnamefont{Edmonds}},
  \bibinfo{author}{\bibfnamefont{R.~P.} \bibnamefont{Campion}},
  \bibinfo{author}{\bibfnamefont{C.~T.} \bibnamefont{Foxon}}, \bibnamefont{and}
  \bibinfo{author}{\bibfnamefont{B.}~\bibnamefont{Gallagher}},
  \bibinfo{journal}{Journ. Appl. Phys.} \textbf{\bibinfo{volume}{95}},
  \bibinfo{pages}{7399} (\bibinfo{year}{2004}).

\bibitem[{\citenamefont{Hrabovsky et~al.}(2002)\citenamefont{Hrabovsky,
  Vanelle, Fert, Yee, Redoules, Sadowski, Kanski, and Ilver}}]{Hrabovsky02}
\bibinfo{author}{\bibfnamefont{D.}~\bibnamefont{Hrabovsky}},
  \bibinfo{author}{\bibfnamefont{E.}~\bibnamefont{Vanelle}},
  \bibinfo{author}{\bibfnamefont{A.~R.} \bibnamefont{Fert}},
  \bibinfo{author}{\bibfnamefont{D.~S.} \bibnamefont{Yee}},
  \bibinfo{author}{\bibfnamefont{J.~P.} \bibnamefont{Redoules}},
  \bibinfo{author}{\bibfnamefont{J.}~\bibnamefont{Sadowski}},
  \bibinfo{author}{\bibfnamefont{J.}~\bibnamefont{Kanski}}, \bibnamefont{and}
  \bibinfo{author}{\bibfnamefont{L.}~\bibnamefont{Ilver}},
  \bibinfo{journal}{Appl. Phys. Lett.} \textbf{\bibinfo{volume}{81}},
  \bibinfo{pages}{2806} (\bibinfo{year}{2002}).

\bibitem[{\citenamefont{Moore et~al.}(2003)\citenamefont{Moore, Ferr\'e,
  Mougin, Moreno, and Daweritz}}]{Moore03}
\bibinfo{author}{\bibfnamefont{G.~P.} \bibnamefont{Moore}},
  \bibinfo{author}{\bibfnamefont{J.}~\bibnamefont{Ferr\'e}},
  \bibinfo{author}{\bibfnamefont{A.}~\bibnamefont{Mougin}},
  \bibinfo{author}{\bibfnamefont{M.}~\bibnamefont{Moreno}}, \bibnamefont{and}
  \bibinfo{author}{\bibfnamefont{L.}~\bibnamefont{Daweritz}},
  \bibinfo{journal}{Journ. Appl. Phys.} \textbf{\bibinfo{volume}{94}},
  \bibinfo{pages}{4530} (\bibinfo{year}{2003}).

\bibitem[{\citenamefont{Lang et~al.}(2005)\citenamefont{Lang, Winter, Pascher,
  Krenn, Liu, and Furdyna}}]{Lang05}
\bibinfo{author}{\bibfnamefont{R.}~\bibnamefont{Lang}},
  \bibinfo{author}{\bibfnamefont{A.}~\bibnamefont{Winter}},
  \bibinfo{author}{\bibfnamefont{H.}~\bibnamefont{Pascher}},
  \bibinfo{author}{\bibfnamefont{H.}~\bibnamefont{Krenn}},
  \bibinfo{author}{\bibfnamefont{X.}~\bibnamefont{Liu}}, \bibnamefont{and}
  \bibinfo{author}{\bibfnamefont{J.~K.} \bibnamefont{Furdyna}},
  \bibinfo{journal}{Phys. Rev. B}
  \textbf{\bibinfo{volume}{72}}, \bibinfo{eid}{024430}
  (\bibinfo{year}{2005}).

\bibitem[{\citenamefont{Beschoten et~al.}(1999)\citenamefont{Beschoten,
  Crowell, Malajovich, Awschalom, Matsukura, Shen, and Ohno}}]{Beschoten99}
\bibinfo{author}{\bibfnamefont{B.}~\bibnamefont{Beschoten}},
  \bibinfo{author}{\bibfnamefont{P.~A.} \bibnamefont{Crowell}},
  \bibinfo{author}{\bibfnamefont{I.}~\bibnamefont{Malajovich}},
  \bibinfo{author}{\bibfnamefont{D.~D.} \bibnamefont{Awschalom}},
  \bibinfo{author}{\bibfnamefont{F.}~\bibnamefont{Matsukura}},
  \bibinfo{author}{\bibfnamefont{A.}~\bibnamefont{Shen}}, \bibnamefont{and}
  \bibinfo{author}{\bibfnamefont{H.}~\bibnamefont{Ohno}},
  \bibinfo{journal}{Physical Review Letters} \textbf{\bibinfo{volume}{83}},
  \bibinfo{pages}{3073} (\bibinfo{year}{1999}).

\bibitem[{\citenamefont{Harmand et~al.}(1989)\citenamefont{Harmand, Matsuno,
  and Inoue}}]{Harmand89}
\bibinfo{author}{\bibfnamefont{J.-C.} \bibnamefont{Harmand}},
  \bibinfo{author}{\bibfnamefont{T.}~\bibnamefont{Matsuno}}, \bibnamefont{and}
  \bibinfo{author}{\bibfnamefont{K.}~\bibnamefont{Inoue}},
  \bibinfo{journal}{Jap. J. Appl. Physics} \textbf{\bibinfo{volume}{28}},
  \bibinfo{pages}{L1101} (\bibinfo{year}{1989}).

\bibitem[{\citenamefont{Liu et~al.}(2005)\citenamefont{Liu, Lim, Titova,
  Dobrowolska, Furdyna, Kutrowski, and Wojtowicz}}]{liu:063904}
\bibinfo{author}{\bibfnamefont{X.}~\bibnamefont{Liu}},
  \bibinfo{author}{\bibfnamefont{W.~L.} \bibnamefont{Lim}},
  \bibinfo{author}{\bibfnamefont{L.~V.} \bibnamefont{Titova}},
  \bibinfo{author}{\bibfnamefont{M.}~\bibnamefont{Dobrowolska}},
  \bibinfo{author}{\bibfnamefont{J.~K.} \bibnamefont{Furdyna}},
  \bibinfo{author}{\bibfnamefont{M.}~\bibnamefont{Kutrowski}},
  \bibnamefont{and}
  \bibinfo{author}{\bibfnamefont{T.}~\bibnamefont{Wojtowicz}},
  \bibinfo{journal}{Journ. Appl. Phys.} \textbf{\bibinfo{volume}{98}},
  \bibinfo{eid}{063904} (\bibinfo{year}{2005}).

\bibitem[{\citenamefont{Primus et~al.}(2005)\citenamefont{Primus, Choi,
  Trampert, Yakunin, Ferr\'e, Wolter, Roy, and Boeck}}]{Primus2005}
\bibinfo{author}{\bibfnamefont{J.-L.} \bibnamefont{Primus}},
  \bibinfo{author}{\bibfnamefont{K.-H.} \bibnamefont{Choi}},
  \bibinfo{author}{\bibfnamefont{A.}~\bibnamefont{Trampert}},
  \bibinfo{author}{\bibfnamefont{M.}~\bibnamefont{Yakunin}},
  \bibinfo{author}{\bibfnamefont{J.}~\bibnamefont{Ferr\'e}},
  \bibinfo{author}{\bibfnamefont{J.~H.} \bibnamefont{Wolter}},
  \bibinfo{author}{\bibfnamefont{W.}~\bibnamefont{Roy}}, \bibnamefont{and}
  \bibinfo{author}{\bibfnamefont{J.~D.} \bibnamefont{Boeck}},
  \bibinfo{journal}{J. Cryst. Growth} \textbf{\bibinfo{volume}{280}},
  \bibinfo{pages}{32} (\bibinfo{year}{2005}).

\bibitem[{\citenamefont{Sinova et~al.}(2004)\citenamefont{Sinova, Jungwirth,
  and Cerne}}]{Sinova04b}
\bibinfo{author}{\bibfnamefont{J.}~\bibnamefont{Sinova}},
  \bibinfo{author}{\bibfnamefont{T.}~\bibnamefont{Jungwirth}},
  \bibnamefont{and} \bibinfo{author}{\bibfnamefont{J.}~\bibnamefont{Cerne}},
  \bibinfo{journal}{Int. Journ. of Modern Phys. B}
  \textbf{\bibinfo{volume}{18}} (\bibinfo{year}{2004}).

\bibitem[{\citenamefont{Berger and Bergmann}(1979)}]{Berger79}
\bibinfo{author}{\bibfnamefont{L.}~\bibnamefont{Berger}} \bibnamefont{and}
  \bibinfo{author}{\bibfnamefont{G.}~\bibnamefont{Bergmann}},
  \emph{\bibinfo{title}{The Hall Effect and its Applications}}
  (\bibinfo{publisher}{Plenum, New York}, \bibinfo{year}{1979}).

\bibitem[{\citenamefont{Jungwirth et~al.}(2002)\citenamefont{Jungwirth, Niu,
  and MacDonald}}]{Jungwirth02}
\bibinfo{author}{\bibfnamefont{T.}~\bibnamefont{Jungwirth}},
  \bibinfo{author}{\bibfnamefont{Q.}~\bibnamefont{Niu}}, \bibnamefont{and}
  \bibinfo{author}{\bibfnamefont{A.~H.} \bibnamefont{MacDonald}},
  \bibinfo{journal}{Phys. Rev. Lett.} \textbf{\bibinfo{volume}{88}},
  \bibinfo{eid}{207208} (\bibinfo{year}{2002}).

\bibitem[{\citenamefont{Ferre}(2001)}]{Ferre2001}
\bibinfo{author}{\bibfnamefont{J.}~\bibnamefont{Ferr\'e}},
  \emph{\bibinfo{title}{Spin dynamics in confined magnetic structures}},
  (\bibinfo{publisher}{Springer, Heidelberg}, \bibinfo{year}{2001}), pp. \bibinfo{pages}{127--160}.

\bibitem[{\citenamefont{Hubert and Sch\"{a}fer}(1998)}]{A.Hubert1198}
\bibinfo{author}{\bibfnamefont{A.}~\bibnamefont{Hubert}} \bibnamefont{and}
  \bibinfo{author}{\bibfnamefont{R.}~\bibnamefont{Sch\"{a}fer}},
  \emph{\bibinfo{title}{Magnetic Domains}} (\bibinfo{publisher}{Springer-Verlag, Berlin}, \bibinfo{year}{1998}).

\bibitem[{\citenamefont{Lemerle et~al.}(1997)\citenamefont{Lemerle, J. Ferr\'{e},
  A.Thiaville, McVitie, and Chappert}}]{Chappert}
\bibinfo{author}{\bibfnamefont{S.}~\bibnamefont{Lemerle}},
  \bibinfo{author}{\bibnamefont{J.Ferre}},
  \bibinfo{author}{\bibnamefont{A.Thiaville}},
  \bibinfo{author}{\bibfnamefont{S.}~\bibnamefont{McVitie}}, \bibnamefont{and}
  \bibinfo{author}{\bibfnamefont{C.}~\bibnamefont{Chappert}},
  \bibinfo{journal}{Kluwer Acad. Publ.} pp. \bibinfo{pages}{537--542}
  (\bibinfo{year}{1997}).

\bibitem[{\citenamefont{Yamanouchi et~al.}(2004)\citenamefont{Yamanouchi,
  Chiba, Matsukura, and Ohno}}]{Yamanouchi04}
\bibinfo{author}{\bibfnamefont{M.}~\bibnamefont{Yamanouchi}},
  \bibinfo{author}{\bibfnamefont{D.}~\bibnamefont{Chiba}},
  \bibinfo{author}{\bibfnamefont{F.}~\bibnamefont{Matsukura}},
  \bibnamefont{and} \bibinfo{author}{\bibfnamefont{H.}~\bibnamefont{Ohno}},
  \bibinfo{journal}{Nature} \textbf{\bibinfo{volume}{429}}
  (\bibinfo{year}{2004}).

\end{thebibliography}

\end{document}